\documentclass[12pt]{article}
\usepackage{graphicx}
\usepackage{graphics}
\usepackage{amsthm}
\usepackage{amssymb, amsmath}
\title{Mellin transforms with only critical zeros:  Legendre functions}
\author{Mark W. Coffey\\
Department of Physics\\
Colorado School of Mines\\
Golden, CO  80401\\
USA\\
mcoffey@mines.edu\\
Matthew C. Lettington\\
School of Mathematics\\
Cardiff University\\
P. O. Box 926\\
Cardiff CP24 4AG\\
UK\\
LettingtonMC@cf.ac.uk}
\date{August 27, 2013}
\begin{document}
\maketitle
\baselineskip=25 pt
\begin{abstract}

We consider the Mellin transforms of certain Legendre functions based upon the ordinary and
associated Legendre polynomials.  We show that the transforms have polynomial factors whose zeros lie all on the critical line Re $s=1/2$.  The polynomials with zeros only on the critical
line are identified in terms of certain $_3F_2(1)$ hypergeometric functions.  
These polynomials possess the functional equation $p_n(s)=(-1)^{\lfloor n/2 \rfloor} p_n(1-s)$. 
Other hypergeometric representations are presented, as well as certain Mellin transforms of  
fractional part and fractional part-integer part functions.
The results should be of interest to special function theory, combinatorial geometry, and 
analytic number theory. 

\end{abstract}
 
\vspace{.25cm}
\baselineskip=15pt
\centerline{\bf Key words and phrases}
\medskip 
Mellin transformation, Legendre polynomial, associated Legendre polynomial, hypergeometric function, critical line, zeros, functional equation

\bigskip
\noindent
{\bf 2010 MSC numbers}
\newline{33C20, 33C45, 42C05, 44A20, 30D05}  

\baselineskip=25pt

\pagebreak
\centerline{\bf Introduction}

\medskip

Mellin transforms are very important in analytic number theory and asymptotic analysis.  They occasionally also find application in signal and image analysis.  In a series of 
investigations, we are determining families of polynomials arising from Mellin transformation
that satisfy the Riemann hypothesis.

In particular, we are considering certain Mellin transforms comprised of classical orthogonal polynomials that yield polynomial factors with zeros only on the critical line Re $s=1/2$ or else only on the real axis.  Such polynomials have many important applications to
analytic number theory, in a sense extending the Riemann hypothesis.  For example,
using the Mellin transforms of Hermite functions, Hermite polynomials multiplied by a
Gaussian factor, Bump and Ng \cite{bumpng} (see also \cite{bumpchoi}) were able to generalize
Riemann's second proof of the functional equation of the zeta function $\zeta(s)$, and to
obtain a new representation for it.  The polynomial factors turn out to be certain 
$_2F_1(2)$ Gauss hypergeometric functions, being certain shifted symmetric Meixner-Pollaczek polynomials \cite{coffeymellin}.  

The polynomials $p_n(x)= ~_2F_1(-n,-x;1;2)=(-1)^n ~_2F_1(-n,x+1;1;2)$ and $q_n(x)=i^n n! p_n(-1/2-ix/2)$ have been studied for combinatorial and number theoretic reasons 
\cite{kirsch,stoll}, and they directly correspond to the Bump and Ng polynomials with $s=-x$.  
We note that these polynomials arise in the counting of the number of lattice points in an
$n$-dimensional octahedron \cite{stoll}.  In fact, combinatorial, geometrical, and coding aspects of $p_n(x)$ at integer argument had been noted in \cite{golomb} and \cite{stanton},
and Lemmas 2.2 and 2.3 of \cite{kirsch} correspond very closely to Lemmas 2 and 3,
respectively, of \cite{stanton}.
For the half-line Mellin transform of Laguerre functions, one may see \cite{coffeymellin}.

The Riemann zeta function arises as the half-line Mellin transform of a theta function, and
of many other functions.  However, these are not the sole type of Mellin transform from which the zeta function may be determined.
Letting $\{x\}$ denote the fractional part of $x$, we quickly review the representation for
Re $s>1$,
$$\int_0^1 \left\{{1 \over t}\right\} t^{s-1}dt={1 \over {s-1}}-{{\zeta(s)} \over s}.$$
For we have, with $\zeta(s,a)$ the Hurwitz zeta function,
$$\int_0^1 \left\{{1 \over t}\right\} t^{s-1}dt=\int_1^\infty \left\{v\right\} v^{-s-1}dv
=\sum_{k=1}^\infty \int_k^{k+1} \left\{v\right\} v^{-s-1}dv$$
$$=\sum_{k=1}^\infty \int_0^1 {v \over {(v+k)^{s+1}}}dv = \int_0^1 v\zeta(s+1,v+1)dv
={1 \over {s-1}}-{{\zeta(s)} \over s}.$$
With $\Gamma$ the Gamma function and $_2F_1$ the Gauss hypergeometric function, in turn we may
generalize this representation to
$$\int_0^1 \left\{{1 \over t}\right\} {t^{s-1} \over {(1-t^b)^\alpha}} dt=
{{\Gamma(1-\alpha)\Gamma\left({{s+b-1} \over b}\right)} \over {\Gamma\left({{s+b-\alpha b-1}
\over b}\right)}}{1 \over {(s-1)}}-{1 \over s}\sum_{j=1}^\infty {1 \over j^s} ~_2F_1\left(\alpha,
{s \over b};1+{s \over b};{1 \over j^b}\right), $$
with $0 \leq \mbox{Re} ~\alpha <1$.  This result uses the Newton series for the Beta function
and the interchange of a double summation:
$$\int_0^1 \left\{{1 \over t}\right\} {t^{s-1} \over {(1-t^b)^\alpha}} dt=
\sum_{\ell=0}^\infty (-1)^\ell {{-\alpha} \choose \ell}\int_0^1 \left\{ {1 \over t} \right\}
t^{s+b\ell-1}dt$$
$$=\sum_{\ell=0}^\infty (-1)^\ell {{-\alpha} \choose \ell}\int_1^\infty \left\{v\right\} v^{-s-b\ell-1}dv =
\sum_{\ell=0}^\infty (-1)^\ell {{-\alpha} \choose \ell}
\sum_{k=1}^\infty \int_0^1 {v \over {(v+k)^{s+b \ell +1}}}dv$$
$$=\sum_{\ell=0}^\infty (-1)^\ell {{-\alpha} \choose \ell}\int_0^1 v\zeta(s+b\ell+1,k+1)dv$$
$$=\sum_{\ell=0}^\infty (-1)^\ell {{-\alpha} \choose \ell}\left[{1 \over {s+b\ell-1}}
-{{\zeta(s+b\ell)} \over {s+b\ell}}\right]$$
$$={{\Gamma(1-\alpha)\Gamma\left({{s+b-1} \over b}\right)} \over {\Gamma\left({{s+b-\alpha b-1}
\over b}\right)}}{1 \over {(s-1)}}-\sum_{\ell=0}^\infty (-1)^\ell {{-\alpha} \choose \ell}
{{\zeta(s+b\ell)} \over {s+b\ell}}$$
$$={{\Gamma(1-\alpha)\Gamma\left({{s+b-1} \over b}\right)} \over {\Gamma\left({{s+b-\alpha b-1}
\over b}\right)}}{1 \over {(s-1)}}-\sum_{\ell=0}^\infty (-1)^\ell {{-\alpha} \choose \ell}
{1 \over {(s+b\ell)}}\sum_{j=1}^\infty {1 \over j^{s+b \ell}}$$
$$={{\Gamma(1-\alpha)\Gamma\left({{s+b-1} \over b}\right)} \over {\Gamma\left({{s+b-\alpha b-1}
\over b}\right)}}{1 \over {(s-1)}}-{1 \over s}\sum_{j=1}^\infty {1 \over j^s} ~_2F_1\left(\alpha,
{s \over b};1+{s \over b};{1 \over j^b}\right). $$
An interesting case occurs for $s=2$, $b=1$ and $\alpha \to 1$:
{\newline \bf Corollary}.
$$\int_0^1 \left\{{1 \over t}\right\}{t \over {(1-t)}}dt=\gamma,$$
where $\gamma$ is the Euler constant.  
{\newline \it Proof}.  In taking the limit, the expression
$${1 \over 2} ~_2F_1(\alpha,2;3;x)={1 \over {(1-\alpha)(2-\alpha)}}\left[1-{{(x-1)(\alpha x-x-1)}
\over {(1-x)^\alpha}}\right]{1 \over x^2},$$
is used, along with the sum $\sum_{j=2}^\infty [1/j+\ln(1-1/j)]=\gamma-1$.  \qed

Of the many Mellin transforms considered in \cite{coffeytapas} (Appendix III) we mention that
$$I_j(s)\equiv \int_0^\infty {x^s \over {(e^x+1)^j}}dx=\Gamma(s+1)\sum_{n=0}^\infty (-1)^n {{(j)_n} \over {n!}}{1 \over {(n+j)^{s+1}}}, ~~~~ \mbox{Re} ~s>-1,$$
and
$$J_j(s)\equiv \int_0^\infty {x^s \over {(e^x-1)^j}}dx=\Gamma(s+1)\sum_{n=0}^\infty {{(j)_n} 
\over {n!}}{1 \over {(n+j)^{s+1}}}, ~~~~ \mbox{Re} ~s>0,$$
may always be written in terms of the zeta function for integers $j \geq 1$.  This follows
since the ratio $(j)_n/n!$ may be reduced and rewritten as a sum of powers of $n+j$.  As
examples we have
$$I_2(s)=\Gamma(s+1)[(1-2^{-s})\zeta(s+1)+(2^{1-s}-1)\zeta(s)],$$
and
$$I_3(s)=\Gamma(s+1)2^{-s-1}[(2^s-4)\zeta(s-1)+3(2-2^s)\zeta(s)+2(2^s-1)\zeta(s+1)].$$
Such integrals are useful for bounding the zeta function on the positive real axis.

In this article, we study the Mellin transforms of certain Legendre functions, and
are able to identify the resulting polynomial factors in terms of certain generalized hypergeometric functions $_3F_2(1)$.  The key result is that these polynomials possess zeros only on the critical line. 

We use standard notation.
Let $_pF_q$ be the generalized hypergeometric function, $(a)_n=\Gamma(a+n)/\Gamma(a)=(-1)^n {{\Gamma(1-a)} \over {\Gamma(1-a-n)}}$ the Pochhammer symbol,
and $B(x,y)=\Gamma(x)\Gamma(y)/\Gamma(x+y)$ the Beta function.  Let $P_\nu^\mu$ be the Legendre functions of the first kind, and $P_n=P_n^0$ the ordinary Legendre polynomials.  Standard
results on orthogonal polynomials may be found in \cite{chihara,szego} and \cite{andrews}
(Chs. 5--7).


We consider Mellin transformations for functions supported on $[0,1]$,
$$({\cal M}f)(s)=\int_0^1 f(x)x^s {{dx} \over x}.  \eqno(1.1)$$
For properties of Mellin transforms, \cite{butzer} and \cite{bleistein} (Ch. 4) may be
consulted.  A related technique is the Master Theorem of Ramanujan, (e.g., \cite{andrews},
section 10.12), relating a Mellin transform to the coefficients of an alternating power series.
We do not rely on this Theorem, but we illustrate it briefly in our context (see the second
proof of Lemma 2).

Put, for Re $s>0$,
$$M_n(s) \equiv \int_0^1 x^{s-1} P_n(x) {{dx} \over \sqrt{1-x^2}}.  \eqno(1.2)$$
We are employing the concept of generalized Mellin transform, for functions for which
the Mellin transform on all of $[0,\infty)$ does not otherwise exist \cite{bleistein} (section 4.3). We note that we could instead consider the Mellin transforms
$$M_n^{(-1)}(s) =\int_0^1 P_n\left({1 \over x}\right){x^{-s} \over \sqrt{1-x^2}}dx,$$
as regards the polynomial factors.  This is because, aside from a phase factor of $i$,
the Gamma factors of this transform are simply the analytic continuation of those of (1.2).
Indeed, the correspondence between (1.2) and $M_n^{(-1)}(s)$ is closely connected with the
functional equation of the polynomial factors.

{\bf Proposition 1}.  For Re $s>0$, 
$$M_n(s)={1 \over \sqrt{\pi}} {{\Gamma\left({{n+s} \over 2}\right)} \over {\Gamma\left( {{n+s+1} \over 2}\right)}} \int_0^{\pi/2} ~_2F_1\left({{1-n} \over 2},-{n \over 2};1-{{(n+s)} \over 2};
\cos^2 \varphi\right) d\varphi. \eqno(1.3)$$

{\bf Proposition 2}.  The polynomial factors $p_n(s)$ of (1.3) satisfy the functional equation $p_n(s)=(-1)^{\lfloor n/2\rfloor}p_n(1-s)$.  Moreover, all of their zeros are on the line Re $s=1/2$.  

In place of the $_3F_2$ representations for $M_n(s)$ of Lemmas 2 and 3, we have the
following representation in terms of a finite sum of $_2F_1(-1)$ functions.
\newline{\bf Proposition 3}.  Let Re $s >0$.  Then  
$$M_n(s)={{(n!)^2} \over 2^n}\Gamma(s)\sum_{k=0}^n {1 \over {(k!)^2}} {{(-1)^k} \over {[(n-k)!]^2}}{{\Gamma(k+1/2)} \over {\Gamma(k+s+1/2)}} ~_2F_1\left({1\over 2}+k-n,s;{1\over 2}+k+s;-1\right).  \eqno(1.4)$$

Generalizing (1.2), for Re $s>0$ we put
$$M_n^m(s) \equiv \int_0^1 x^{s-1} P_n^m(x) {{dx} \over \sqrt{1-x^2}}.  \eqno(1.5)$$
We recall that for associated Legendre functions $P_\nu^m(x)=(-1)^m(1-x^2)^{m/2} \left({d \over
{dx}}\right)^m P_\nu(x)$ and thus $P_n^m(x)=0$ for $m>n$.  We will take $m$ to be nonnegative
in the following.  Otherwise, for negative index the following relation (\cite{grad}, p. 1008)
could be employed:
$$P_\nu^{-m}(x)=(-1)^m {{\Gamma(\nu-m+1)} \over {\Gamma(\nu+m+1)}}P_\nu^m(x).$$

{\bf Proposition 4}.  The polynomial factors of $M_n^m(s)$ satisfy the functional equation $p_n^m(s)=(-1)^{\lfloor n/2 \rfloor} p_n^m(1-s)$ and moreover have all of their zeros on
the critical line. 

The following section of the paper contains the proof of these Propositions.  Section
3 contains various supporting and reference Lemmas.  Some of these Lemmas present 
results of special function theory that may be of interest in themselves.
We note that one of us (MWC) has presented a subset of these results in an AMS special 
session \cite{mwcams}.  The final Discussion section mentions a connection of the polynomial 
factors of the Mellin transforms with continuous Hahn polynomials.

We note that there is a possibility to physically realize these Mellin transforms in an
all-optical system.  A one or two dimensional Mellin transform may be carried out by
making a hologram of a logarithmically scaled function, then optically Fourier
transforming the reconstructed wavefront.  The Fourier transform may be accomplished 
simply with a lens.  The more challenging aspect is the logarithmic scale change,
performable with a spatial light modulator.  Much of the design of such processing has been
given \cite{wu}. 

\medskip
\centerline{\bf Proof of Propositions} 
\medskip

{\it Proposition 1}.  We provide two proofs for the integral representation.
Method 1.  We use Lemma 3(c), along with
$${{\left({1 \over 2}\right)_j} \over {(1)_j}}={1 \over \pi}B\left({1 \over 2},j+{1 \over 2}\right)={1 \over \sqrt{\pi}} {{\Gamma(j+1/2)} \over {j!}}={1 \over \pi}\int_0^\pi \cos^{2j}
\varphi ~d\varphi.$$
Then
$$ ~_3F_2\left({{1-n} \over 2},-{n \over 2},{1 \over 2};1,1-{{(n+s)} \over 2};1\right)
=\sum_{j \geq 0} {{\left({{1-n} \over 2}\right)_j \left(-{n \over 2}\right)_j \left({1 \over 2}\right)_j} \over {\left(1-{{n+s} \over 2}\right)_j (1)_j}} {1 \over {j!}}$$
$$={1 \over \pi} \sum_{j \geq 0} {{\left({{1-n} \over 2}\right)_j \left(-{n \over 2}\right)_j}  \over {\left(1-{{n+s} \over 2}\right)_j}} {1 \over {j!}}\int_0^\pi \cos^{2j}\varphi ~d\varphi.$$
Now
$\int_0^\pi \cos^m \varphi ~d\varphi=0$ for $m$ odd.  Therefore, we may write
$$ ~_3F_2\left({{1-n} \over 2},-{n \over 2},{1 \over 2};1,1-{{(n+s)} \over 2};1\right)
={1 \over \pi} \sum_{j \geq 0} {{\left({{1-n} \over 2}\right)_{j/2} \left(-{n \over 2}\right)_{j/2}}  \over {\left(1-{{n+s} \over 2}\right)_{j/2}}} {1 \over {(j/2)!}}\int_0^\pi \cos^j \varphi ~d\varphi$$
$$={1 \over \pi} \int_0^\pi ~_2F_1\left({{1-n} \over 2},-{n \over 2};1-{{(n+s)} \over 2};
\cos^2 \varphi\right) ~d\varphi.$$
From the evenness of the integrand on $[0,\pi]$, (2.6) follows.  \qed

Method 2.  We may use the Beta transform \cite{grad} (p. 850)
$$\int_0^1 (1-x)^{-1/2}x^{-1/2} ~_2F_1 \left({{1-n} \over 2},-{n \over 2};1-{{n+s} \over 2};x
\right)dx = \pi ~_3F_2\left({{1-n} \over 2},-{n \over 2},{1 \over 2};1,1-{{(n+s)} \over 2};1\right).  $$
Then making the change of variable $x=\cos^2 \varphi$, we again obtain the Proposition. \qed
  

{\it Proposition 2}.  Up to $s$-{\it independent} factors and choice of normalization, by
Proposition 1 we may take
$$p_n(s)={{\Gamma\left({{n+s} \over 2}\right)} \over {\Gamma\left({{s+\epsilon} \over 2}\right)}}
\int_0^{\pi/2} ~_2F_1\left({{1-n} \over 2},-{n \over 2};1-{{(n+s)} \over 2};\cos^2 \varphi \right)
~d\varphi,$$
where $\epsilon=0$ for $n$ even and $=1$ for $n$ odd.
By \cite{grad} (p. 1043, \# 9.131.2) we transform the argument of the $_2F_1$ function to
$1-\cos^2 \varphi=\sin^2 \varphi$.  Since one of $1/\Gamma(-n/2)$ or $1/\Gamma[(1-n)/2]$ will be
zero depending upon whether $n$ is even or odd, respectively, we have
$$p_n(s)={{\Gamma\left(1-{{(n+s)} \over 2}\right)} \over {\Gamma\left(1-{s \over 2}\right)}}
{{\Gamma\left({{1+n-s} \over 2}\right)} \over {\Gamma\left({{1-s} \over 2}\right)}}{{\Gamma\left({{n+s} \over 2}\right)} \over {\Gamma\left({{s+\epsilon} \over 2}\right)}}
\int_0^{\pi/2} ~_2F_1\left({{1-n} \over 2},-{n \over 2};{{(1-n+s)} \over 2};\sin^2 \varphi \right)
~d\varphi.$$
Since
$$\int_0^{\pi/2} \sin^{2k}\varphi ~d\varphi=\int_0^{\pi/2} \cos^{2k}\varphi ~d\varphi
={1 \over 2}B\left({1 \over 2},k+{1 \over 2}\right),$$
and using the functional equation of the Gamma function, we may equally well write
$$p_n(s)={\pi \over {\Gamma\left(1-{s \over 2}\right)}}
{{\Gamma\left({{1+n-s} \over 2}\right)} \over {\Gamma\left({{1-s} \over 2}\right)}}{1 \over {\sin\pi\left({{n+s} \over 2}\right)\Gamma\left({{s+\epsilon} \over 2}\right)}}
\int_0^{\pi/2} ~_2F_1\left({{1-n} \over 2},-{n \over 2};{{(1-n+s)} \over 2};\cos^2 \varphi \right)
~d\varphi.$$
When $n$ is even, $\epsilon=0$, 
$$\Gamma\left({s \over 2}\right)\Gamma\left(1-{s \over 2}\right)={\pi \over {\sin\pi(s/2)}},$$
leaving the denominator factor $\Gamma\left({{1-s} \over 2}\right)$.
When $n$ is odd, $\epsilon=1$, 
$$\Gamma\left({{s+1} \over 2}\right)\Gamma\left({{1-s} \over 2}\right)={\pi \over {\cos\pi(s/2)}},$$
leaving the denominator factor $\Gamma\left(1-{s \over 2}\right)$.
Hence the factor $(-1)^{\lfloor n/2 \rfloor}$ emerges as $\sin (\pi s/2)/\sin [\pi(n+s)/2] =(-1)^{n/2}$ when $n$ is even and as $\cos (\pi s/2)/\sin [\pi(n+s)/2]=(-1)^{(n-1)/2}$ when 
$n$ is odd, and the functional equation of $p_n(s)$ follows.  

In order to show that $p_n(s)$ has zeros only on the critical line, we first establish
the difference equation
$$[n^2+n-1+4s-2s(s+1)]p_n(s)\left({{s+\epsilon} \over 2}-1\right)\left({{s+n+1} \over 2}\right)
p_n(s)$$
$$+[(s+2)(s+3)-(n^2+n-2)-4(s+2)]\left({{s+\epsilon} \over 2}-1\right)\left({{s+\epsilon} \over
2}\right)p_n(s+2)$$
$$+(s-1)(s-2)\left({{s+n+1} \over 2}\right)\left({{s+n-1} \over 2}\right)p_n(s-2)=0,$$
with $\epsilon=0$ for $n$ even and $=1$ for $n$ odd.  We use the ordinary differential
equation satisfied by Legendre polynomials,
$$(1-x^2)P_n''(x)-2xP_n'(x)+n(n+1)P_n(x)=0.$$
Putting $P_n(x)=\sqrt{1-x^2}f(x)$, we find
$${1 \over \sqrt{1-x^2}}\left[(n^2+n-1-(n^2+n-2)x^2)f(x)+4x(x^2-1)f'(x)+(x^2-1)^2f''(x)
\right]=0.$$
We then integrate the quantity in square brackets by parts and use the definition
$M_n(s)=\int_0^1x^{s-1}f(x)dx$.  There results the difference equation for the Mellin
transforms
$$[n^2+n-1+4s-2s(s+1)]M_n(s)+[(s+2)(s+3)-(n^2+n-2)-4(s+2)]M_n(s+2)$$
$$+(s-1)(s-2)M_n(s-2)=0.$$
The Mellin transforms are of the form
$$M_n^m(s)={\sqrt{\pi} \over 2^{\lfloor n+1 \rfloor}} {{\Gamma\left ({{s+\epsilon} \over 2} \right)} \over {\Gamma\left({s \over 2}+{{n+1} \over 2}\right)}}p_n(s),$$
so that by repeatedly using the functional equation $\Gamma(z+1)=z\Gamma(z)$, the
difference equation for $p_n(s)$ is found.

Now using shifted polynomials $q(s)=p_n(s+1/2)$ and putting $s \to s+1/2$, we have the
difference equation
$$\left[n^2+n+1+4s-2\left(s+{1 \over 2}\right)\left(s+{3 \over 2}\right)\right]\left(
{{s+n+1} \over 2}+{1 \over 4}\right)\left({{s+\epsilon} \over 2}+{1 \over 4}\right)q(s)$$
$$+\left[\left(s+{5 \over 2}\right)\left(s+{7 \over 2}\right)-(n^2+n-2)-4\left(s+{5 \over 2}\right)\right]\left({{s+\epsilon} \over 2}-{3 \over 4}\right)\left({{s+\epsilon} \over 2}+{1 \over 4}\right)q(s+2)$$
$$+\left(s-{1 \over 2}\right)\left(s-{3 \over 2}\right)\left({{s+n} \over 2}+{3 \over 4}\right)
\left({{s+n} \over 2}-{1 \over 4}\right)q(s-2)=0.$$
We note the factorization
$$\left(s+{5 \over 2}\right)\left(s+{7 \over 2}\right)-(n^2+n-2)-4\left(s+{5 \over 2}\right)=\left(s-n+{1 \over 2}\right)\left(s+n+{3 \over 2}\right).$$
It then follows that if $r_k$ is a root of $q$, $q(r_k)=0$, that
$$\left(r_k+\epsilon-{3 \over 2}\right)\left(r_k+\epsilon +{1 \over 2}\right)\left(r_k-n+{1
\over 2}\right)q(r_k+2)$$
$$=-\left(r_k-{1 \over 2}\right)\left(r_k-{3 \over 2}\right)
\left(r_k+n-{1 \over 2}\right)q(r_k-2).$$
When $n$ is even, a factor of $r_k-3/2$ cancels on both sides, and
when $n$ is odd, a factor of $r_k-1/2$ cancels on both sides.  In either case,
equality of the absolute value of both sides provides a necessary condition that
Re $r_i=0$ for all the zeros of $q$.  Hence the zeros of $p_n(s)$ lie on the
critical line.  \qed

{\it Proposition 3}.  We have the following generating function,
$$~_0F_1\left(-;1;{1 \over 2}t(x-1)\right)~_0F_1\left(-;1;{1 \over 2}t(x+1)\right)
=I_0\left(\sqrt{2}\sqrt{t(x-1)}\right)I_0\left(\sqrt{2}\sqrt{t(x+1)}\right)$$
$$=\sum_{n=0}^\infty {{P_n(x)t^n} \over {(n!)^2}}, $$
giving
$$\sum_{n=0}^\infty {t^n \over {(n!)^2}}M_n(s)=\int_0^1 {x^{s-1} \over \sqrt{1-x^2}}
I_0\left(\sqrt{2}\sqrt{t(x-1)}\right)I_0\left(\sqrt{2}\sqrt{t(x+1)}\right)dx.$$
We now insert the power series for $I_0$ \cite{grad} (p. 961), so that
$$\sum_{n=0}^\infty {t^n \over {(n!)^2}}M_n(s)=\sum_{k=0}^\infty \sum_{\ell=0}^\infty
{t^{k+\ell} \over {2^k2^\ell (k!)^2(\ell!)^2}} \int_0^1 {x^{s-1} \over \sqrt{1-x^2}}
(x-1)^k (x+1)^\ell dx$$
$$=\sum_{m=0}^\infty {t^m \over 2^m} \sum_{k=0}^m {1 \over {(k!)^2}}{1 \over {[(m-k)!]^2}}
\int_0^1 {x^{s-1} \over \sqrt{1-x^2}} (x-1)^k (x+1)^{m-k} dx.  \eqno(2.1)$$
The integral evaluates in terms of the Gauss hypergeometric function,
$$-i\int_0^1 x^{s-1}(x+1)^{m-k-1/2}(x-1)^{k-1/2}dx=(-1)^k {{\Gamma(k+1/2)} \over {\Gamma(k+s+1/2)}} ~_2F_1\left({1\over 2}+k-m,s;{1\over 2}+k+s;-1\right).$$
Then reading off the coefficient of $t^n$ on both sides of (2.1) yields the Proposition.
\qed

{\it Remarks}.  Proposition 3 is consistent with the relations
$$~_2F_1\left(s,{1 \over 2};s+{1 \over 2};-1\right)={\sqrt{\pi} \over 2^s}{{\Gamma\left(s+{1 \over 2}\right)} \over {\Gamma^2\left({{s+1} \over 2}\right)}}, \eqno(2.2a)$$
$$~_2F_1\left(s,-{1 \over 2};s+{1 \over 2};-1\right)={\sqrt{\pi} \over 2^s}\Gamma\left(s+{1 \over 2}\right)\left[{1 \over {\Gamma^2\left({{s+1} \over 2}\right)}}+{s \over 2}{1 \over {\Gamma^2\left({s \over 2}+1\right)}}\right], \eqno(2.2b)$$
and
$$~_2F_1\left(s,{1 \over 2};s+{3 \over 2};-1\right)=-{\sqrt{\pi} \over 2^{s-1}}\Gamma\left(s+{3 \over 2}\right)\left[{2 \over s}{1 \over {\Gamma^2\left({s \over 2}\right)}}-{1 \over {\Gamma^2\left({{s+1} \over 2}\right)}}\right]. \eqno(2.2c)$$
(2.2a) is Kummer's identity, and the others may be obtained with the aid of contiguous relations.

The $_2F_1(-1)$ function of (1.4) may be transformed to functions $_2F_1(1/2)$.

{\it Proposition 4}.  We have (\cite{grad}, p. 1015)
$$P_n^m(x)=(-1)^m {{(2n-1)!!} \over {(n-m)!}} (1-x^2)^{m/2} x^{n-m} ~_2F_1\left({{n-m} \over 2},
{{m-n+1} \over 2};{1 \over 2}-n;{1 \over x^2}\right),$$
giving
$$M_n^m(s)=i^{m-n}{\sqrt{\pi} \over 2} \Gamma\left({{m+1} \over 2}\right)\Gamma\left({1 \over 2}
-n\right) \left\{ {{\Gamma\left({s \over 2}\right) ~_3F_2\left({{m-n} \over 2},{{m+n+1} \over 2},
{s \over 2};{1 \over 2}, {{1+m+s} \over 2};1\right)} \over {\Gamma\left({{1-m-n} \over 2}\right)
\Gamma\left({{1+m-n} \over 2}\right)\Gamma\left({{1+m-s} \over 2}\right)}} \right.$$
$$\left. +{{2i\Gamma\left({{s+1} \over 2}\right) ~_3F_2\left({{1+m-n} \over 2},1+{{m+n} \over 2},
{{s+1} \over 2};{3 \over 2}, 1+{{m+s} \over 2};1\right)} \over {\Gamma\left({{m-n} \over 2}\right)
\Gamma\left(-{{(m+n)} \over 2}\right)\Gamma\left(1+{{m+s} \over 2}\right)}} \right \}.$$
The first line on the right side of this expression gives $M_n^m$ for $m-n$ even, and the
second line for $m-n$ odd.  The key step in demonstrating the functional equation is
transforming the $_3F_2$ functions so that a numerator parameter is twice a denominator parameter.
Then the Beta transformation 
$$\int_0^1 (1-x)^{\beta-1}x^{\beta-1} ~_2F_1\left({{1-n} \over 2},-{n \over 2};1-{{(n+s)} \over 2};x\right)dx$$
$$ =2^{1-2\beta} {{\sqrt{\pi} \Gamma(\beta)} \over {\Gamma(\beta+1/2)}} ~_3F_2\left(\beta,{{1-n} \over 2},-{n \over 2};2\beta,1-{{(n+s)} \over 2};1\right),$$
together with a transformation of the argument of the $_2F_1$ function from $x$ to $1-x$ may be
used to verify the functional equation.  Each of the three transformations given in the Appendix
may used to put the $_3F_2$ functions of $M_n^m(s)$ in suitable form.  We illustrate this 
procedure using (A.3) and the $_3F_2$ function for $m-n$ even,
$$~_3F_2\left({{m-n} \over 2},{{m+n+1} \over 2},{s \over 2};{1 \over 2}, {{1+m+s} \over 2};1 \right)= {{\Gamma\left({{1+m-n} \over 2}\right)\Gamma\left(1+{{m-n} \over 2}\right)\Gamma\left(
{{s+1} \over 2}\right)\Gamma\left({{1+m+s} \over 2}\right)} \over {\sqrt{\pi} \Gamma(1+m-n)
\Gamma\left({{1+m-n+s} \over 2}\right)\Gamma\left({{1+n+s} \over 2}\right)}}$$
$$\times ~_3F_2\left({{m+1} \over 2},{{m-n} \over 2}{{1+m-n} \over 2};m+1,{{1+m+s} \over 2};1\right).$$
There is only one $_3F_2$ function on the right side as the other has a vanishing $1/\Gamma[(m-n)/2]$ prefactor.  
The functional equation then follows. 

The demonstration that all of the zeros of $p_n^m(s)$ lie on the critical line follows
very closely that corresponding to the proof of Proposition 2.  The ordinary differential
equation satisfied by associated Legendre polynomials,
$$(1-x^2)P_n^{m''}(x)-2xP_n^{m'}(x)+\left[n(n+1)-{m^2 \over {1-x^2}}\right]P_n^m(x)=0,$$
leads to the difference equation of the Mellin transforms,
$$[n^2-m^2+n-1+4s-2s(s+1)]M_n(s)+[(s+2)(s+3)-(n^2+n-2)-4(s+2)]M_n(s+2)$$
$$+(s-1)(s-2)M_n(s-2)=0.$$
So in the difference equation for $p_n^m(s)$ and the corresponding shifted polynomial
$q(s)=p_n^m(s+1/2)$, only the coefficient of $p_n^m(s)$ or $q(s)$ is modified, thus
having no effect on the rest of the proof.  It follows once again that the zeros of $q(s)$
are pure imaginary and thus the zeros of $p_n^m(s)$ all lie on the critical line.  \qed

\medskip
\centerline{\bf Lemmas}  
\medskip

{\bf Lemma 1}.  Let $n \geq 1$ and Re $s>0$.  Then
$$M_n(s)={1 \over n}[(2n-1)M_{n-1}(s+1)-(n-1)M_{n-2}(s)],  \eqno(3.1)$$
and 
$$M_0(s)={\sqrt{\pi} \over 2}{{\Gamma\left({s \over 2}\right)} \over
{\Gamma\left({{s+1} \over 2}\right)}}, ~~~~~~
M_1(s)={\sqrt{\pi} \over 2}{{\Gamma\left({{s+1} \over 2}\right)} \over
{\Gamma\left({{s} \over 2}+1\right)}}.$$
Notice that with $M_{-1}(s)$ arbitrary but finite, the recursion (3.1) properly
degenerates to $M_1(s)=M_0(s+1)$.  It is easily seen that $M_0(s)=B(s/2,1/2)/2$, 
where $B$ is the Beta function.

{\it Proof}.  The mixed recurrence (3.1) follows from  
$$(n+1)P_{n+1}(x)=(2n+1)xP_n(x)-nP_{n-1}(x).$$
The expressions for $M_0$ and $M_1$ follow from the integral representation
$$\int_0^1 x^{a-1}(1-x^2)^{b-1}dx={1 \over 2}B\left({a \over 2},b\right), ~~~~\mbox{Re} ~a>0,
~~\mbox{Re} ~b>0.$$
\qed

There are a great many transformations of $_3F_2(z)$ and $_3F_2(1)$ functions, including
Thomae's.  Rather than list the result of several of these applied to the Mellin transforms
$M_n(s)$, we present the result of obtaining these transforms directly through integration.

{\bf Lemma 2}.  Let $(2n+1)!! \equiv (2n+1)(2n-1)\cdots 3$.  
The following representations for $M_n(s)$ hold for Re $s>0$ when $n$ is
even and for Re $s>-1$ when $n$ is odd.  
(a)
$$M_{2n+1}(s)={{(-1)^n} \over 2} {{\left({{2-s} \over 2}\right)_n} \over {\left({{s+1} \over
2}\right)_{n+1}}} ~_3F_2\left({1 \over 2},{{s+1} \over 2},{s \over 2}; {s \over 2}-n,
{s \over 2}+n+{3 \over 2};1\right),$$
(b)
$$M_{2n}(s)={{(-1)^n} \over 2} {{\left({{1-s} \over 2}\right)_n} \over {\left({s \over
2}\right)_{n+1}}} ~_3F_2\left({1 \over 2},{{s+1} \over 2},{s \over 2}; {s \over 2}-n+{1 \over 2},{s \over 2}+n+1;1\right),$$
(c)
$$M_{2n}(s)=(-1)^n {{(2n-1)!!} \over {2^{n+1}n!}}{{\sqrt{\pi}\Gamma\left({s \over 2}\right)} \over
{\Gamma\left({{s+1} \over 2}\right)}} ~_3F_2\left(-n,n+{1 \over 2},{s \over 2};{1 \over 2},{{s+1}
\over 2};1\right),$$
(d)
$$M_{2n+1}(s)=(-1)^n {{(2n+1)!!} \over {2^nn!}}{{\sqrt{\pi}\Gamma\left({{s+1} \over 2}\right)} \over {s\Gamma\left({s \over 2}\right)}} ~_3F_2\left(-n,n+{3 \over 2},{{s+1} \over 2};{3 \over 2},{s \over 2}+1;1\right),$$
and (e)
$$M_n(s)=(-1)^n {{(2n-1)!!} \over {2^{n+1}n!}}{{\pi\Gamma\left({{s-n} \over 2}\right)} \over
{\Gamma\left({{1-n} \over 2}\right)\Gamma\left({{s-n+1} \over 2}\right)}} ~_3F_2\left(-n,{1 \over 2},{{s-n} \over 2};{{1-n} \over 2},{{s-n+1} \over 2};1\right).$$

{\it Proof}.  (a) and (b) follow by binomially expanding $(1-x^2)^{-1/2}=\sum_{\ell=0}^\infty
{{-1/2} \choose \ell} (-1)^\ell x^{2\ell}$ in the integrand of $M_n(s)$ and performing 
term-by-term integration.  
For the rest we recall the representation \cite{grad} (p. 850) for Re $s>0$ and Re $\nu >1$
$$\int_0^1 x^{s-1}(1-x^2)^\nu ~_2F_1(-n,a;b;x^2)dx={1 \over 2}B\left(\nu+1,{s \over 2}\right)
~_3F_2\left(-n,a,{s \over 2};b,\nu+1+{s \over 2};1\right).$$
Then we employ various $_2F_1$ expressions for $P_n(x)$ from \cite{grad} (p. 1025).
In particular, for (e), we use
$$P_n(x)={{(2n-1)!!} \over {n!}}x^n ~_2F_1\left(-{n \over 2},{{1-n} \over 2};{1 \over 2}-n;{1
\over x^2}\right),$$
along with a simple change of variable.  \qed

We also present as an example use of Ramanujan's Master Theorem an alternative proof of part (c),
using the hypergeometric expression
$$P_{2n}(x)=(-1)^n {{(2n-1)!!} \over {2^n n!}} ~_2F_1\left(-n,n+{1 \over 2};{1 \over 2};x^2
\right).$$
We then obtain the power series
$${{P_{2n}(x)} \over \sqrt{1-x^2}}=(-1)^n {{(2n-1)!!} \over {2^n n!}}\sum_{m=0}^\infty \sum_{j=0}^m {{(-n)_j(n+1/2)_j} \over {(1/2)_j}} {{(-1)^{m-j}} \over {j!}} {{-1/2} \choose {m-j}} x^{2m}.$$
Writing $\Phi(x^2)=\sum_{k=0}^\infty (-1)^k \phi(k)x^{2k}$, we have
$$\phi(m)=(-1)^n {{(2n-1)!!} \over {2^n n!}} \sum_{j=0}
^m {{(-n)_j(n+1/2)_j} \over {(1/2)_j}} {{(-1)^j} \over {j!}} {{-1/2} \choose {m-j}},$$
giving
$${\pi \over {\sin \pi s}}\phi(-s)=(-1)^n {{(2n-1)!!} \over {2^n n!}}{\pi \over {\sin \pi s}}
{{\sqrt{\pi} ~_3F_2\left(-n,n+{1 \over 2},s;{1 \over 2},s+{1 \over 2};1\right)} \over
{\Gamma(1-s)\Gamma(s+1/2)}}$$
$$=(-1)^n {{(2n-1)!!} \over {2^n n!}} \sqrt{\pi}{{\Gamma(s)} \over {\Gamma(s+1/2)}} ~_3F_2\left(-n,n+{1 \over 2},s;{1 \over 2},s+{1 \over 2};1\right).$$
Taking into account the change of variable
$\int_0^1 u^{s-1}\Phi(u)du=2\int_0^1 x^{2s-1} \Phi(x^2)dx$, so that replacing $s \to s/2$ just
above, we obtain agreement with $M_{2n}(s)$ in part (c).  \qed

{\bf Lemma 3}.  Alternative hypergeometric representation for $M_n(s)$.
(a)
$$M_n(s)={1 \over 2^{n+1}}\sum_{k=0}^{\lfloor n/2 \rfloor} {{(-1)^k (2n-2k)!} \over 
{k!(n-k)!(n-2k)!}} B\left({{s+n} \over 2}-k,{1 \over 2}\right), $$
(b)
$$M_n(s)=2^{n-1}{{\Gamma\left(n+{1 \over 2}\right)\Gamma\left({{n+s} \over 2}\right)} \over
{n!\Gamma\left({{n+s+1} \over 2}\right)}} ~_3F_2\left({{1-n} \over 2},-{n \over 2},{{1-n-s}
\over 2};{1 \over 2}-n,1-{n \over 2}-{s \over 2};1\right),$$
and (c)
$$M_n(s)={\sqrt{\pi} \over 2} {{\Gamma\left({{n+s} \over 2}\right)} \over
{\Gamma\left({{n+s+1} \over 2}\right)}} ~_3F_2\left({{1-n} \over 2},-{n \over 2},{1 \over 2};
1,1-{{(n+s)} \over 2};1\right).$$
The forms (b) and (c) are convenient for presenting the truncation at degree $\lfloor 
n/2 \rfloor$, due to the presence of both the numerator parameters $-n/2$ and $(1-n)/2$.

{\it Proof}.  Parts (a) and (b) follow from a series in \cite{grad} (p. 1025),
$$P_n(x)={1 \over 2^n}\sum_{k=0}^{\lfloor n/2 \rfloor} {{(-1)^k (2n-2k)!} \over {k!(n-k)!(n-2k)!}}x^{n-2k}. $$
Relations such as
$${{\Gamma\left({{s+n+1} \over 2}\right)} \over {\Gamma\left({{s+n+1} \over 2}-k\right)}}
=(-1)^k\left({{1-n-s} \over 2}\right)_k,$$
and
$${{\Gamma\left({{s+n} \over 2}\right)} \over {\Gamma\left({{s+n} \over 2}-k\right)}}
=(-1)^k\left(1-{{(n+s)} \over 2}\right)_k,$$
are then used.   
Part (c) follows from using Laplace's integral (3.4).  \qed

{\it Remarks}.
Various transformation corollaries follow from the equalities of the expressions in Lemmas 2 and 3.

For $s=1$, the $_3F_2$ function in Lemma 3(b) reduces to particular $_2F_1$ values,
$$_2F_1\left(-{n \over 2},-{n \over 2};{1 \over 2}-n;1\right)={{\sqrt{\pi}\Gamma\left({1 \over
2}-n\right)} \over {\Gamma^2 \left({{1-n} \over 2}\right)}}.$$
This gives the special values  
$$M_n(1)={{(-1)^n \pi^2} \over {2\Gamma^2 \left({{1-n} \over 2}\right)\Gamma^2 \left({n \over 2}+1\right)}}.$$  

Here is another way to calculate the transforms $M_n(s)$.  Use the Fourier transform
$$P_n(x)={{(-i)^n} \over \sqrt{2\pi}}\int_{-\infty}^\infty t^{-1/2}J_{n+1/2}(t)e^{itx}dt,
~~~~|x|<1,$$  
where $J_n$ is the Bessel function of the first kind of order $n$.  Then interchange integrations. 

Here is $M_n(s)$ written as a $1/2$-line transform:
$$M_n(s)=\int_0^\infty \tanh^{s-1} u {{P_n(\tanh u)} \over {\cosh u}}du,$$
wherein the weight function $1/\cosh u$ is self-reciprocal, up to scaling, on the 
full line, under the (exponential) Fourier transform.

We also have
$$M_n(s)=\int_0^{\pi/2} \cos^{s-1} \theta P_n(\cos \theta)d\theta. $$
Using \cite{grad} (p. 1027) allows alternative $_3F_2$ forms to be obtained, including
$$M_n(s)={\sqrt{\pi} \over 2} {{\Gamma\left({{n+s} \over 2}\right)} \over {\Gamma\left({{n+s+1} \over 2}\right)}} ~_3F_2\left({1 \over 2},{{1-n} \over 2},-{n \over 2};
1,1-{{(n+s)} \over 2};1\right),$$
in agreement with Lemma 3(c).

{\bf Lemma 4}.  (Generating function of Mellin transforms.)  For Re $s>0$,
$$G(t,s) \equiv \sum_{k=0}^\infty M_k(s)t^k =\int_0^1 {1 \over {(1-x^2)^{1/2}}}{x^{s-1} \over 
{(1-2tx+t^2)^{1/2}}}dx$$
$$={\sqrt{\pi} \over {(1+t^2)^{1/2}}}\left\{{{\Gamma\left({s \over 2}\right)} \over {2\Gamma\left({{s+1} \over 2}\right)}} ~_3F_2\left[{1 \over 4},{3 \over 4},{s \over 2};{1 \over 2};
{{s+1} \over 2};{{4t^2} \over {(1+t^2)^2}}\right] \right.$$
$$\left. + {t \over {(1+t^2)}}{{\Gamma\left({{s+1} \over 2}\right)} \over {\Gamma\left({s \over 2}\right)}} ~_3F_2\left[{3 \over 4},{5 \over 4},{{s+1} \over 2};{3 \over 2},{s \over 2}+1; {{4t^2} \over {(1+t^2)^2}}\right] \right\}.$$
The first line of the right member yields $M_{2k}(s)$ and the second line, $M_{2k+1}(s)$.

{\it Proof}.  A generating function of Legendre polynomials is
$$\sum_{k=0}^\infty t^k P_k(x)={1 \over \sqrt{1-2tx+t^2}}.$$
Then binomially expanding we have
$$\int_0^1 {1 \over {(1-x^2)^{1/2}}}{x^{s-1} \over {(1-2tx+t^2)^{1/2}}}dx
=\sum_{\ell=0}^\infty {{-{1 \over 2}} \choose \ell}(-1)^\ell \int_0^1 {x^{2\ell+s-1} \over
{(1-2tx +t^2)^{1/2}}}dx$$
$$={1 \over {(1+t^2)^{1/2}}}\sum_{\ell=0}^\infty {{-{1 \over 2}} \choose \ell}{{(-1)^\ell } \over {(2\ell+s)}} ~_2F_1\left({1 \over 2},2\ell+s;1+2\ell+s;{{2t} \over {1+t^2}}\right).$$
We then interchange sums, and separate terms of even and odd summation index.  \qed

{\it Remarks}.  Another generating function relation is
$$\sum_{k=0}^\infty {t^k \over {k!}}M_k(s)=\int_0^1 e^{xt}x^{s-1} {{J_0(t\sqrt{1-x^2})}
\over \sqrt{1-x^2}}dx$$
$$=\int_0^1 e^{\sqrt{1-u^2}t}(1-u^2)^{s/2-1} J_0(tu)du.$$
The first equality in
$$\sum_{n=0}^\infty {{P_n(x)} \over {n!}}r^n = e^{xr}J_0(r\sqrt{1-x^2})
=e^{xr} e^{-r^2/4}\sum_{n=0}^\infty {r^{2n} \over {4^n n!}}L_n(1-x^2)$$
follows from Laplace's integral for $P_n(x)$.  Here $L_n$ is the Laguerre polynomial of degree $n$.

{\bf Lemma 5}.
{\newline The polynomials $p_n$ satisfy the following recursion relation, with $p_0=p_1=1$.}
{\newline For $n$ even,}
$$p_n(s)={2 \over n}[(2n-1)sp_{n-1}(s+1)-(n-1)(s+n-1)p_{n-2}(s)],$$
and for $n$ odd,
$$p_n(s)={1 \over n}[(2n-1)p_{n-1}(s+1)-2(n-1)(s+n-1)p_{n-2}(s)].$$

{\it Proof}.  By Proposition 1 or the proof of Proposition 2, the Mellin transforms are of the 
form 
$$M_n(s)={\sqrt{\pi} \over 2^n} {{p_n(s) \Gamma\left({{s+\epsilon} \over 2} \right)} 
\over {\Gamma\left({s \over 2}+{{n+1} \over 2}\right)}},$$
where $\epsilon=0$ for $n$ even and $=1$ for $n$ odd.  
(The $2^{-n}$ normalization makes
$p_n(s)$ of leading coefficient $2^{\lfloor n/2 \rfloor}$.)  
Then using Lemma 1 and the functional equation of the Gamma function gives the result.  \qed

{\bf Lemma 6}.  Let Re $s>0$.  Then
$$M_n^m(s)={1 \over {(n-m)}}[(2n-1)M_{n-1}^m(s+1)-(n+m-1)M_{n-2}^m(s)],$$
and 
$$M_0^m(s)=\delta_{m0}{\sqrt{\pi} \over 2}{{\Gamma\left({s \over 2}\right)} \over
{\Gamma\left({{s+1} \over 2}\right)}}, ~~~~~~
M_1^m(s)=\delta_{m0}{\sqrt{\pi} \over 2}{{\Gamma\left({{s+1} \over 2}\right)} \over
{\Gamma\left({{s} \over 2}+1\right)}}-\delta_{m1}{1 \over s},$$
where $\delta_{nm}$ is the Kronecker symbol.  Furthermore,
$M_{m+1}^m(s)=(2m+1)M_m^m(s+1)$ and 
$$M_m^m(s)=(-1)^m(2m-1)!! {{\Gamma\left({s \over 2}\right)\Gamma\left({{m+1} \over 2}\right)}
\over {2\Gamma\left({{s+m+1} \over 2}\right)}}.$$

{\it Proof}.  This follows from
$$(2\nu+1)xP_\nu^m(x)=(\nu-m+1)P_{\nu+1}^m(x)+(\nu+m)P_{\nu-1}^m(x),$$
$P_0^m(x)=\delta_{0m}$, $P_1^m(x)=\delta_{0m}x-\delta_{1m}\sqrt{1-x^2}$,
$P_m^m(x)=(-1)^m (2m-1)!!(1-x^2)^{m/2}$, and $P_{m+1}^m(x)=(2m+1)xP_m^m(x)$.  \qed

{\it Remarks}.
A way to calculate these Mellin transforms, especially for odd values of $m$, is via the relation (e.g., \cite{grad}, p. 1008)
$$P_\nu^m(x)=(-1)^m (1-x^2)^{m/2} {d^m \over {dx^m}}P_\nu(x).$$

Using $P_n(1)=1$ and \cite{nbs} (p. 338), integrating by parts we have
$$M_n^1(s)={\sqrt{\pi} \over 2^{s-1}}{{\Gamma(s)} \over {\Gamma\left({{s-n} \over 2}\right)
\Gamma\left({{s+n+1} \over 2}\right)}}-1. \eqno(3.2a)$$
As $M_0^1=0$, it is evident that this expression recovers Legendre's duplication formula
when $n=0$.  I.e., alternatively we may write
$$M_n^1(s)={{\Gamma\left({s \over 2}\right)\Gamma\left({{s+1} \over 2}\right)} \over 
{\Gamma\left({{s-n} \over 2}\right)\Gamma\left({{s+n+1} \over 2}\right)}}-1. \eqno(3.2b)$$
In so far as
$$\Gamma\left({{s-n} \over 2}\right)\Gamma\left({{s+n+1} \over 2}\right)=\sqrt{\pi}2^{n-s}
\Gamma(s-n)(s-n+1)\left({{3+s-n} \over 2}\right)_{n-1},$$
$M_n^1(s)$ is a rational function of $s$.  More generally, when $m$ is odd, $M_n^m(s)$ is a
rational function of $s$.

{\bf Lemma 7}.  For $m$ an even integer,
for $n$ even,
$$p_n^m(s)={2 \over {n-m}}[(2n-1)sp_{n-1}^m(s+1)-(n+m-1)(s+n-1)p_{n-2}^m(s)],$$
and for $n$ odd,
$$p_n^m(s)={1 \over {n-m}}[(2n-1)p_{n-1}^m(s+1)-2(n+m-1)(s+n-1)p_{n-2}^m(s)].$$
Further, $p_m^m=(-1)^m(2m-1)!!(m-1)!!$ and $p_{m+1}^m=(2m+1)p_m^m$. 

{\it Proof}.  For $m$ even, the Mellin transforms are of the form 
$$M_n^m(s)={\sqrt{\pi} \over 2^{\lfloor n-m/2+1 \rfloor}} {{p_n^m(s) \Gamma\left
({{s+\varepsilon} \over 2} \right)} \over {\Gamma\left({s \over 2}+{{n+1} \over 2}\right)}}, $$
where $\epsilon=0$ for $n$ even and $=1$ for $n$ odd.  For $n<m$, they vanish.
Then by using Lemma 6 and the functional equation of the Gamma function gives the recurrences.
\qed

{\bf Lemma 8}.  Let $p \equiv (n+1)(n+2)\cdots (n+m)$.  Then
$$M_n^m(s)={{p \pi} \over {2\Gamma\left({{n+s+1}\over 2}\right)}}\left\{\sqrt{\pi}
\Gamma\left({{n+s}\over 2}\right) {{~_4F_3\left({1 \over 2},1,{{1-n} \over 2},-{n \over 2};
1-{m \over 2},1+{m \over 2},1-{{(n+s)} \over 2};1\right)} \over {\Gamma\left({{1-m} \over 2}
\right)\Gamma(1-m)\Gamma\left({{1+m} \over 2}\right)m!}} \right.$$
$$-2in \left.\Gamma\left({{n+s-1}\over 2}\right) {{~_4F_3\left(1,1,{{1-n} \over 2},1-{n \over 
2};{{3-m} \over 2},{{3+m} \over 2},{{3-(n+s)} \over 2};1\right)} \over {\Gamma(2-m)
\Gamma\left(-{m \over 2}\right)\Gamma\left({m \over 2}\right)(m+1)!}} \right\}. \eqno(3.3)$$

{\it Proof}.  Laplace's integral for $P_n^m$ is \cite{grad} (p. 1001)
$$P_n^m(x)={1 \over \pi}(n+1)(n+2)\cdots (n+m)\int_0^\pi [x+\sqrt{x^2-1}\cos \theta]^n
\cos m\theta ~d\theta. \eqno(3.4)$$
We binomially expand the integrand to develop $M_n^m(s)$,
$$M_n^m(s)=-p\sum_{\ell=0}^n {n \choose \ell}i \int_0^\pi \int_0^1 x^{n-\ell+s-1} (x^2-1)^{(
\ell-1)/2} \cos m\theta \cos^\ell \theta dx d\theta$$
$$={p\over 2}\sum_{\ell=0}^n {n \choose \ell}i^\ell {{\Gamma\left({{\ell+1}\over 2}\right)
\Gamma\left({{n+s-\ell}\over 2}\right)} \over {\Gamma\left({{n+s+1}\over 2}\right)}}
\int_0^\pi \cos m\theta \cos^\ell \theta ~d\theta $$
$$={p\over 2}\sum_{\ell=0}^n {n \choose \ell}i^\ell {{\Gamma\left({{\ell+1}\over 2}\right)
\Gamma\left({{n+s-\ell}\over 2}\right)} \over {\Gamma\left({{n+s+1}\over 2}\right)}}
{{2^\ell \pi^2 (-1)^\ell \ell!} \over {(\ell-m)!(\ell+m)!\Gamma\left({{1-\ell-m} \over 2}
\right)\Gamma\left({{1-\ell+m} \over 2}\right)}}.$$  
The trigonometric integral here is given in \cite{grad} (p. 374).  \qed

{\it Remarks}.
(3.3) may be written in several ways using functional equations of the Gamma function.
It includes many special cases, such as for $n=m$, and $m=1$.  For the latter case,
there is reduction to a $_2F_1(1)$ function that gives (3.2).  

Following up the example Mellin transforms of the Introduction, we have the following.
We let $[x]$ denote the integer part of $x$.
\newline{\bf Lemma 9}.  (a) For Re $s>1$,
$$\int_0^1 \left\{{1 \over t}\right\}\left[{1 \over t}\right] t^{s-1}dt={1 \over {s(s-1)}}
\left\{s-1-s\zeta(s-1)+2\zeta(s-1)+(s-1)[\zeta(s)-1]\right\},$$
with the value at $s=2$ being $[\zeta(2)-1]/2$, (b) for Re $s>2$,
$$\int_0^1 \left\{{1 \over t}\right\}\left[{1 \over t}\right]^2 t^{s-1}dt={1 \over {s(s-1)}}
\left[(3-s)\zeta(s-2)+(2s-3)\zeta(s-1)+(1-s)\zeta(s)\right],$$
with the value at $s=3$ being $\zeta(2)/2-\zeta(3)/3-1/6$, and (c) for $n \geq 1$ an integer
and Re $s>n$,
$$\int_0^1 \left\{{1 \over t}\right\}\left[{1 \over t}\right]^n t^{s-1}dt={1 \over {s(s-1)}}
\left\{-\sum_{\ell=0}^{n-1} (-1)^{n-\ell}\left[{n \choose \ell}-{n \choose {\ell+1}}(s-1)\right]
\zeta(s-\ell-1)\right.$$
$$\left.+(-1)^{n+1}(s-1)\zeta(s)\right\}.$$  For this last integral, the value at $s=n+1$ is
given by
$$\int_0^1 \left\{{1 \over t}\right\}\left[{1 \over t}\right]^n t^n dt=
{1 \over {n(n+1)}}\left\{-\sum_{\ell=0}^{n-2}(-1)^{n-\ell}\left[{n \choose \ell}-{n \choose {\ell+1}}n\right]\zeta(n-\ell)-1 \right.$$
$$\left. +(-1)^{n+1}n\zeta(n+1)\right\}.$$
(d) For Re $s>n+1$,
$$\int_0^1 \left\{{1 \over t}\right\}^2\left[{1 \over t}\right]^n t^{s-1}dt=-{1 \over {s(s-1)(s-2)}}\left\{\sum_{\ell=0}^{n-1} (-1)^{n-\ell}{n \choose \ell}\left[2\zeta(s-\ell-2) \right. \right.$$ 
$$\left. \left. +2(s-2)\zeta(s-\ell-1)+(s-1)(s-2)\zeta(s-\ell)\right]
+2(s-2)\zeta(s-n-1)+(s-1)(s-2)\zeta(s-n)\right\}.$$
The value for $s\to n+2$ is given by
$$-{1 \over {n(n+1)(n+2)}}\left\{\sum_{\ell=0}^{n-2}(-1)^{n-\ell}{n \choose \ell}[2\zeta(n-\ell)
+2n\zeta(n-\ell+1)+n(n+1)\zeta(n-\ell+2)]\right.$$
$$\left. +2-n(n-1)\zeta(2)-n^2(n+1)\zeta(3)\right\}.$$
(e)  Let $H_n=\sum_{k=1}^n {1 \over k}$ be the $n$th harmonic number and $\psi^{(j)}(z)$
be the polygamma function (e.g. \cite{nbs,grad}).  Let $s>1$ be an integer.  Then
$$\int_0^1 \left\{{1 \over x}\right\}\left\{{1 \over {1-x}}\right\}x^{s-1}dx
=\sum_{\ell=1}^s {{(-1)^\ell} \over {\ell!}}\psi^{(\ell-1)}(1)+{1 \over 2}-H_s+
+{{1-s2^{s-1}} \over {s2^s}}$$
$$+{1 \over {s(s-1)}}\sum_{k=1}^\infty \left[{k^s \over {(k+1)^{s-3}}}+{{[s-(k+1)^2]} \over
{(k+2)^s}}(k+1)^{s-1}\right],$$
where $\psi^{(\ell-1)}(1)=(-1)^\ell(\ell-1)!\zeta(\ell)$.

{\it Proof}.  We first consider convergence, of the more general integrals
$$\int_0^1 \left\{{1 \over t}\right\}^\alpha \left[{1 \over t}\right]^\beta t^{s-1}dt
=\int_1^\infty \{v\}^\alpha [v]^\beta v^{-s-1}dv$$
$$=\sum_{k=1}^\infty k^\beta \int_k^{k+1} {{(v-k)^\alpha} \over v^{s+1}}dv
=\sum_{k=1}^\infty k^\beta \int_0^1 {u^\alpha \over {(u+k)^{s+1}}}du.$$
Noting that
$${1 \over {(k+1)^{s+1}}}<{1 \over {(u+k)^{s+1}}}<{1 \over k^{s+1}},$$
we have
$${1 \over {(\alpha+1)}}\sum_{k=1}^\infty {k^\beta \over {(k+1)^{s+1}}}
<\int_0^1 \left\{{1 \over t}\right\}^\alpha \left[{1 \over t}\right]^\beta t^{s-1}dt
<{1 \over {(\alpha+1)}}\sum_{k=1}^\infty {k^\beta \over k^{s+1}}.$$
Therefore, for convergence of the integral we require Re $s >\beta$.  This is regardless of
the value of $\alpha$, as long as Re $\alpha>-1$.

Next, we will repeatedly use the relation $\sum_{k=1}^\infty {{(k+1)^p} \over {(k+1)^s}}
=\zeta(s-p)-1$ for Re ~$s>p+1$.  (a) We write
$$\int_0^1 \left\{{1 \over t}\right\}\left[{1 \over t}\right] t^{s-1}dt
=\int_1^\infty \{v\}[v]v^{s-1}dv=\sum_{k=1}^\infty \int_k^{k+1} \{v\}[v]v^{s-1}dv$$
$$=\sum_{k=1}^\infty k \int_0^1 {u \over {(u+k)^{s+1}}}du$$
$$={1 \over {s(s-1)}}\sum_{k=1}^\infty \left[{1 \over k^{s-2}} -{{k(s+k)} \over {(k+1)^s}}\right]$$
$$={1 \over {s(s-1)}}\left\{[1-\zeta(s-1)](s-1)+(s-1)[\zeta(s)-1]+\zeta(s-1)\right\}$$
$$={1 \over {s(s-1)}}\left\{s-1-s\zeta(s-1)+2\zeta(s-1)+(s-1)[\zeta(s)-1]\right\}.$$
For the case of $s \to 2$, we note that 
$$-\lim_{s\to 2}(s-2)\zeta(s-1)=-\lim_{s\to 1}(s-1)\zeta(s)=-1.$$
(b) goes similarly.  For (c) we use
$$\int_0^1 \left\{{1 \over t}\right\}\left[{1 \over t}\right]^n t^{s-1}dt
={1 \over {s(s-1)}}\sum_{k=1}^\infty \left[{1 \over k^{s-n-1}}-{{k^n(k+s)} \over {(k+1)^s}}
\right]$$
$$={1 \over {s(s-1)}}\left[\zeta(s-n-1)-\sum_{k=1}^n {{\sum_{\ell=0}^n (-1)^{n-\ell}{n \choose 
\ell}(k+1)^\ell[(k+1)+(s-1)]} \over {(k+1)^s}}\right]$$
$$={1 \over {s(s-1)}}\left\{\zeta(s-n-1)-\sum_{\ell=0}^n\left[(-1)^{n-\ell}{n \choose \ell}
[\zeta(s-\ell-1)]+(s-1)[\zeta(s-\ell)-1]\right] \right\}.$$
Herein the $\zeta(s-n-1)$ term is cancelled by a $\ell=n$ term of the sum.  Since
$\sum_{\ell=0}^n (-1)^{n-\ell} {n \choose \ell}=(1-1)^n=0$, with a shift of index in the
summation with the $(s-1)$ factor, the stated result for the integral follows. 
For the value at $s=n+1$, the sum has $(n-s+1)\zeta(s-n)$ for the $\ell=n-1$ term.
Since $\lim_{s\to n+1}(n-s+1)\zeta(s-n)=-\lim_{s\to 1}(s-1)\zeta(s)=-1$ the value for $s=n+1$
obtains.  
(d) goes similarly as (c), with 
$$\int_0^1 \left\{{1 \over t}\right\}^2\left[{1 \over t}\right]^n t^{s-1}dt
={1 \over {s(s-1)(s-2)}}\sum_{k=1}^\infty k^n\left[{2 \over k^{s-2}}-{{(2k^2+2ks+s^2-s)} \over
{(k+1)^s}}\right].$$

For (e) we employ a sublemma.  In addition, at times we use the functional equation of the polygamma function,
$$\psi^{(\ell)}(u+2)=\psi^{(\ell)}(u+1)+{{(-1)^\ell \ell!} \over {(u+1)^\ell}}.$$

{\bf Sublemma}.  For $n\geq 1$ an integer, let
$${\cal S}_n(u) \equiv \sum_{k=2}^\infty {1 \over {(u+k)^{n+1}}}{1 \over {(u+k-1)}}.$$
Then
$${\cal S}_j(u)=\sum_{\ell=0}^j {{(-1)^\ell} \over {\ell!}} \psi^{(\ell)}(u+2)-\psi(u+1)$$
$$=\sum_{\ell=1}^j {{(-1)^\ell} \over {\ell!}}\left[\psi^{(\ell)}(u+1)+{{(-1)^\ell \ell!}
\over {(u+1)^{\ell+1}}}\right]+{1 \over {u+1}},$$
wherein the functional equation of $\psi^{(\ell)}(u+2)$ has been used.

{\it Proof}.  Starting with the sum
$$\sum_{k=2}^\infty {1 \over {(u+k)}}{1 \over {(w+k-1)}}={1 \over {(u-w+1)}}\left[\psi(u+2)-
\psi(w+1)\right],$$
we differentiate $j$ times with respect to $u$, resulting in
$$(-1)^j j!\sum_{k=2}^\infty {1 \over {(u+k)^{j+1}}}{1 \over {(w+k-1)}}=\sum_{\ell=0}^j
{{(-1)^{j-\ell}(j-\ell)!{\ell \choose j}} \over {(u-w+1)^{\ell-j+1}}}\psi^{(\ell)}(u+2)
-{{(-1)^j j!} \over {(u-w+1)^{j+1}}}\psi(w+1).$$
We then cancel a factor of $(-1)^j$ on both sides, put $w$ to $u$, and rearrange.  \qed

Now
$$\int_0^1 \left\{{1 \over x}\right\}\left\{{1 \over {1-x}}\right\}x^{s-1}dx
=\int_1^\infty {{\{t\}} \over t^2}t^{1-s}\left\{ {t \over {t-1}} \right\} dt$$
$$=\int_1^2 {{\{t\}} \over t^{s+1}}\left\{ {t \over {t-1}} \right\} dt
+\sum_{k=2}^\infty \int_k^{k+1} {{(t-k)} \over t^{s+1}}\left\{ {t \over {t-1}} \right\} dt.
\eqno(3.5)$$
The latter sum becomes
$$\sum_{k=2}^\infty \int_k^{k+1} {{(t-k)} \over t^{s+1}}\left\{ {t \over {t-1}} \right\} dt
=\sum_{k=2}^\infty \int_k^{k+1} {{(t-k)} \over t^{s+1}}\left({t \over {t-1}}-1\right)dt$$
$$=\sum_{k=2}^\infty \int_k^{k+1} {{(t-k)} \over t^{s+1}}{{dt} \over {(t-1)}}$$
$$=\sum_{k=2}^\infty \int_0^1 {u \over {(u+k)^{s+1}}}{{du} \over {(u+k-1)}}.$$
Thus, we require the integral
$$\sum_{k=2}^\infty \int_0^1 {u \over {(u+k)^{s+1}}}{{du} \over {(u+k-1)}}=\int_0^1 u
{\cal S}_s(u)du,$$
to be evaluated according to the Sublemma.  Using two elementary integrals performed by
integrating by parts, $\int_0^1 {u \over {u+1}}du=1-\ln 2$ and $\int_0^1 {u \over {(u+1)^2}}du=-{1 \over 2}+\ln 2$, and further integrating by parts, we find
$$\int_0^1 u{\cal S}_j(u)du=\sum_{\ell=1}^j {{(-1)^\ell} \over {\ell!}}\left[\psi^{(\ell-2)}
(1)-\psi^{(\ell-2)}(2)+\psi^{(\ell-1)}(1)\right]+{1 \over 2}$$
$$+\sum_{\ell=2}^j 2^{-\ell}{{(2^\ell-\ell-1)} \over {\ell(\ell-1)}}. \eqno(3.6)$$
The latter sum is given by
$$\sum_{\ell=2}^j 2^{-\ell}{{(2^\ell-\ell-1)} \over {\ell(\ell-1)}}={{j-1} \over j}-\sum_{\ell=2}^j
{{\ell+1} \over {2^\ell \ell (\ell-1)}}$$
$$={{j-1} \over j}-\sum_{\ell=2}^j {1\over 2^\ell}\left({2 \over {\ell-1}}-{1 \over \ell}\right)
={{j-1} \over j}+{{(1-j 2^{j-1})} \over {j 2^j}}.$$
   
For the other contribution in (3.5), we have 
$$\int_1^2 {{\{t\}} \over t^{s+1}}\left\{ {t \over {t-1}} \right\} dt
=\int_0^1 {u \over {(u+1)^{s+1}}}\left\{{{u+1} \over u}\right\}du
=\int_0^1 {u \over {(u+1)^{s+1}}}\left\{{1 \over u}\right\}du$$
$$=\int_1^\infty{{\{t\}} \over t^3}{t^{s+1} \over {(1+t)^{s+1}}}dt
=\sum_{k=1}^\infty \int_k^{k+1} {{(t-k)t^{s-2}} \over {(1+t)^{s+1}}}dt$$
$$=\sum_{k=1}^\infty \int_0^1 {{u(u+k)^{s-2}} \over {(u+k+1)^{s+1}}}du$$
$$={1 \over {s(s-1)}}\sum_{k=1}^\infty \left[{k^s \over {(k+1)^{s-3}}}+{{[s-(k+1)^2]} \over
{(k+2)^s}}(k+1)^{s-1}\right].$$
Combining this sum with (3.6), using the functional equation of the polygamma function
and the definition of the harmonic numbers gives the final result.  \qed

{\it Remarks}.  The Sublemma may also be proved using the following approach.
We observe that
$${\cal S}_n(u)=\sum_{k=2}^\infty \left[{1 \over {(u+k)^n}}{1 \over {(u+k-1)}}-{1 \over
{(u+k)^{n+1}}}\right].$$
Hence ${\cal S}_n={\cal S}_{n-1}-\zeta(n+1,u+2)$, where
$\zeta(n+1,x)=(-1)^{n+1}\psi^{(n)}(x)/n!$.  It follows that
$${\cal S}_n(u)=f(u)-\sum_{k=0}^{n-1} \zeta(k+2,u+2),$$ with
$${\cal S}_0(u)={1 \over {1+u}}, ~~~~~~{\cal S}_1(u)={1 \over {1+u}}-\psi'(u+2)={{2+u} \over {(1+u)^2}}-\psi'(u+1).$$
Thus ${\cal S}_0(u)$ suffices to determine $f(u)$, and
$${\cal S}_n(u)={1 \over {1+u}}-\sum_{k=0}^{n-1} \zeta(k+2,u+2).$$

For a given integer $s$ value in (e), the infinite sum also evaluates in terms of $\zeta(n)$
values.

For $s=1$, the sum and integral below (3.6) are
$$\sum_{k=1}^\infty \int_0^1 {{u(u+k)^{-1}} \over {(u+k+1)^2}}du
=\sum_{k=1}^\infty \int_0^1\left[{{k+1} \over {(u+k+1)^2}}+{k \over {u+k+1}}-{k \over {u+k}}
\right]du$$
$$=\sum_{k=1}^\infty {1 \over {(k+2)}}[1+k(k+2)(\ln(k+2)-2\ln(k+1)+\ln k)]$$
$$=-{1 \over 2}+\gamma.$$
With $-\psi(1)=\gamma$ and $1/2-H_1=-1/2$, it follows that $\int_0^1 \left\{{1 \over x}\right\}\left\{{1 \over {1-x}}\right\}dx=2\gamma-1$.

\medskip
\centerline{\bf Discussion}  
\medskip

Within the Askey scheme of hypergeometric polynomials, the continuous Hahn polynomials occur
at the $_3F_2(1)$ level.  Our Mellin transforms are closely connected with instances of these
polynomials, which are given by (e.g., \cite{andrews} p. 331, \cite{askey})
$$p_n(x;a,b,c,d)=i^n {{(a+c)_n(a+d)_n} \over {n!}} ~_3F_2\left(-n,n+a+b+c+d-1,a+ix;a+c,a+d;1
\right).$$
For example, for the transform $M_{2n}(s)$, its polynomial factors are proportional to
$p_n\left(-{{is} \over 2};-n,0,{1 \over 2},{1 \over 2}-n\right)$.  
The continuous Hahn polynomials are orthogonal on the line with respect to the measure
$${1 \over {2\pi}}\Gamma(a+ix)\Gamma(b+ix)\Gamma(c-ix)\Gamma(d-ix)dx.$$
Due to the Parseval relation for the Mellin transform, 
$$\int_0^\infty f(x)g^*(x)dx={1 \over {2\pi i}}\int_{(0)} ({\cal M}f)(s) ({\cal M}g)^*(s)ds,$$
the polynomial factors $p_n(1/2+it)$ form an orthogonal family with respect to a suitable
measure with $\Gamma$ factors.  Since orthogonal polynomials have real zeros, this approach
provides another way of showing that $p_n(s)$ has zeros only on the critical line.


The Gegenbauer polynomials $C_n^\lambda(x)$ are orthogonal on $[-1,1]$ with weight function
$(1-x^2)^{\lambda-1/2}$.  The associated Legendre polynomials are related to them via
$$C_{n-m}^{m+1/2}(x)={1 \over {(2m-1)!!}} {{d^mP_n(x)} \over {dx^m}}=(-1)^m 2^m {{m!}\over 
{(2m)!}}(1-x^2)^{-m/2} P_n^m(x).$$
Elsewhere \cite{coffeylettunpub} we develop a suitable generalization for the Mellin transforms of Gegenbauer functions.  This would provide another approach for proving Proposition 4.


We expect that our work will have connections with the counting of lattice points in polytopes,
thus with combinatorial geometry, a polytope being a region described by a set of linear
inequalities.  In this context, the Ehrhart polynomial \cite{beck} counts lattice points, and
it has a functional equation.  Along with the Ehrhart polynomial one may associate a
Poincar\'{e} series, of the form $P(t)=U(t)/(1-t)^n$, with $U$ a polynomial such that
$U(1) \neq 0$.  An example form of Poincar\'{e} series is 
$$P(t)={{\prod_{j=1}^k (1+t+\ldots+t^{n_j})} \over {(1-t)^n}},$$
wherein $n_1,\ldots,n_k$ are positive integers.  We expect that other various Ehrhart polynomials
are of hypergeometric form, and have all of their zeros on a line.

\pagebreak
\centerline{\bf Appendix:  Selected transformations of $_3F_2(1)$}

\medskip
The following three transformations are valuable in the proof of Proposition 4.
$$~_3F_2(a,b,c;d,e;1) = {{\Gamma(e-a-b)\Gamma(e)} \over {\Gamma(e-a)\Gamma(e-b)}}
~_3F_2(a,b,d-c;d,1+a+b-e;1)$$
$$-{{\Gamma(a+b-e)\Gamma(d)\Gamma(e)\Gamma(d+e-a-b-c)} \over {\Gamma(a)\Gamma(b)\Gamma(d-c)
\Gamma(d+e-a-b)}} ~_3F_2(e-a,e-b,d+e-a-b-c;1+e-a-b,d+e-a-b;1).  \eqno(A.1)$$
$$~_3F_2(a,b,c;d,e;1) = {{\Gamma(1+a-d)\Gamma(1+b-d)\Gamma(1+c-d)\Gamma(d)\Gamma(e)} \over
{\Gamma(a)\Gamma(b)\Gamma(c)\Gamma(1+e-d,2-d)}}$$
$$\times ~_3F_2(1+a-d,1+b-d,1+c-d;1+e-d,2-d;1)$$
$$+ {{\Gamma(1+a-d)\Gamma(1+c-d)} \over {\Gamma(1-d)\Gamma(1+a+c-d)}} ~_3F_2(a,c,e-b;1+a+c-d,e;1).
\eqno(A.2)$$
$$~_3F_2(a,b,c;d,e;1) = {{\Gamma(1+a-d)\Gamma(1+b-d)\Gamma(1+c-d)\Gamma(d)\Gamma(e)} \over
{\Gamma(a)\Gamma(b)\Gamma(c)\Gamma(1+e-d,2-d)}}$$
$$\times ~_3F_2(1+a-d,1+b-d,1+c-d;1+e-d,2-d;1)$$
$$+ {{\Gamma(1+a-d)\Gamma(1+b-d)\Gamma(1+c-d)\Gamma(e)} \over {\Gamma(1-d)\Gamma(1+a+b-d)\Gamma(1+a+c-d)\Gamma(e-a)}}$$
$$\times ~_3F_2(a,1+a-d,1+a+b+c-d-e;1+a+b-d,1+a+c-d;1).  \eqno(A.3)$$

\pagebreak


\begin{thebibliography}{99}
\bibitem{nbs}M. Abramowitz and I. A. Stegun,
{Handbook of Mathematical Functions, Washington, National Bureau of Standards (1964).}
\bibitem{andrews}G. E. Andrews, R. Askey, and R. Roy, 
{Special Functions, Cambridge University Press (1999).}
\bibitem{askey}R. Askey,
{Continuous Hahn polynomials, J. Phys. A {\bf 18}, L1017-L1019 (1985).}
\bibitem{beck}M. Beck and S. Robins,
{Computing the continuous discretely:  Integer-point enumeration in polyhedra, Springer (2007).}
\bibitem{bleistein}N. Bleistein and R. A. Handelsman,
{Asymptotic expansions of integrals, Dover (1986).}
\bibitem{bumpchoi}D. Bump, K.-K. Choi, P. Kurlberg, and J. Vaaler,
{A local Riemann hypothesis, I, Math. Z. {\bf 233}, 1-19 (2000).}
\bibitem{bumpng}D. Bump and E. K.-S. Ng,
{On Riemann's zeta function, Math. Z. {\bf 192}, 195-204 (1986).}
\bibitem{butzer}P. Butzer and S. Jansche,
{A direct approach to the Mellin transform, J. Fourier Analysis Appls. {\bf 3}, 325-376 
(1997).}
\bibitem{chihara}T. Chihara,
{An introduction to orthogonal polynomials, Gordon and Breach (1978).}
\bibitem{coffeymellin}M. W. Coffey,
{Special functions and the Mellin transforms of Laguerre and Hermite functions, 
Analysis {\bf 27}, 95-108 (2007).}  
\bibitem{coffeyxi}M. W. Coffey,
{Theta and Riemann xi function representations from harmonic oscillator eigenfunctions,
Phys. Lett. A {\bf 362}, 352-356 (2007).}
\bibitem{coffeytapas}M. W. Coffey,
{Conjecturing the optimal order of the components of the Li/Keiper constants, Contemp. Math.
{\bf 457}, 135-159 (2008), Tapas in Experimental Mathematics, eds. T. Amdeberhan and V. H. Moll,
American Mathematical Society.}
\bibitem{coffeylettunpub}M. W. Coffey and M. C. Lettington,
{unpublished (2013).}
\bibitem{mwcams}M. W. Coffey,
{AMS western sectional meeting, Special session on Special functions, combinatorics, and
analysis, Tucson, AZ (2012).}
\bibitem{golomb}S. W. Golomb and L. R. Welch,
{Perfect codes in the Lee metric and the packing of polyominoes, SIAM J. Appl. Math.
{\bf 18}, 302-317 (1970).}
\bibitem{grad}I. S. Gradshteyn and I. M. Ryzhik,
{Table of Integrals, Series, and Products, Academic Press, New York (1980).}
\bibitem{kirsch}P. Kirschenhofer, A. Peth\"{o}, and R. F. Tichy,
{On analytical and diophantine properties of a family of counting polynomials, 
Acta Sci. Math. (Szeged) {\bf 65}, 47-59 (1999).}
\bibitem{stanton}R. G. Stanton and D. D. Cowan,
{Note on a ``square" functional equation, SIAM Rev. {\bf 12}, 277-279 (1970).}
\bibitem{stoll}T. Stoll and R. F. Tichy,
{Diophantine equations involving general Meixner and Krawtchouk polynomials, Quaest. Math.
{\bf 28}, 105-115 (2005).}
\bibitem{szego}G. Szeg\"{o},
{Orthogonal Polynomials, Vol. 23 of AMS colloquium Publications, American Mathematical
Society, Providence, RI (1975).}
\bibitem{wu}C. Y. Wu, A. R. D. Somervell, T. G. Haskell, and T. H. Barnes,
{Optical Mellin transform through Haar wavelet transformation, Optics Comm. {\bf 227}, 75-82 (2003).}
\end{thebibliography}
\end{document}